\documentclass[sigconf]{acmart}

\AtBeginDocument{
  }

\setcopyright{none}

\author{Shiyang Chen\footnote{This work is done during his internship at Amazon}}
\affiliation{%
  \institution{Rutgers University}
  \country{}
}
\email{shiyang.chen@rutgers.edu}

\author{Xiang Song}
\affiliation{%
  \institution{Amazon}
  \country{}
}
\email{xiangsx@amazon.com}

\author{Vasiloudis Theodore}
\affiliation{%
  \institution{Amazon}
  \country{}
}
\email{thvasilo@amazon.com}

\author{Hang Liu}
\affiliation{%
  \institution{Rutgers University}
  \country{}
}
\email{hl1097@scarletmail.rutgers.edu}

\usepackage{balance}
\usepackage{microtype}
\usepackage{graphicx}
\usepackage{amsmath}
\usepackage{xurl}
\usepackage{multirow}
\usepackage{mathtools}
\usepackage{booktabs}
\usepackage{tikz}
\usepackage{soul}
\usepackage{url}
\usepackage{hyperref}
\usepackage{subfig}
\usepackage[normalem]{ulem}
\usepackage{sidecap}
\usepackage{xcolor}

\newcommand{\name}{\textsc{Deal}}

\newcommand*\circled[1]{\tikz[baseline=(char.base)]{
            \node[shape=circle,fill,inner sep=1pt] (char) {\footnotesize \textcolor{white}{#1}};}}

\begin{document}

\title{{\name}: Distributed End-to-End GNN Inference for All Nodes}
\date{}

\begin{abstract}

Graph Neural Networks (GNNs) are a new research frontier with various applications and successes. The end-to-end inference for all nodes, is common for GNN embedding models, which are widely adopted in applications like recommendation and advertising. While sharing opportunities arise in GNN tasks (i.e., inference for a few nodes and training), the potential for sharing in full graph end-to-end inference is largely underutilized because traditional efforts fail to fully extract sharing benefits due to overwhelming overheads or excessive memory usage. 

This paper introduces {\name}, a distributed GNN inference system that is dedicated to end-to-end inference for all nodes for graphs with multi-billion edges. First, we unveil and exploit an untapped sharing opportunity during sampling, and maximize the benefits from sharing during subsequent GNN computation. Second, we introduce memory-saving and communication-efficient distributed primitives for lightweight 1-D graph and feature tensor collaborative partitioning-based distributed inference. Third, we introduce partitioned, pipelined communication and fusing feature preparation with the first GNN primitive for end-to-end inference. 
With {\name}, the end-to-end inference time on real-world benchmark datasets is reduced up to $7.70\times$ and the graph construction time is reduced up to $21.05\times$, compared to the state-of-the-art.
 
\end{abstract}

\maketitle

\pagestyle{plain}

\section{Introduction}

Graph learning is emerging as a cutting-edge area in machine learning research~\cite{zheng2022distributed,wang2023efficient, wu2022graph,wu2023certified,chen2023bitgnn,zhu2021graph,jiang2020graph,polisetty2023gsplit,min2022graph,adiletta2023characterizing,yao2024fedgcn,besta2023high,chen2023tango}. Researchers from a wide range of areas, i.e., computer vision~\cite{han2022vision}, natural language processing~\cite{lin2021bertgcn,orogat2023maestro,colas2022gap}, cybersecurity~\cite{he2022illuminati,shaham2020enhancing}, chemistry~\cite{wang2023motif}, material science~\cite{banerjee20013}, bioinformatics~\cite{zhang2021graph} are exploring the possibility of leveraging GNNs to tackle their problems~\cite{hoang2023protecting,cuza2022spatio,peng2021graph,liu2022aspect, hope2021gddr,choma2018graph,dutta2023power,besta2022motif}. The recent successes in such tasks (e.g., chip design~\cite{zhang2019circuit}, computational fluid dynamics~\cite{lino2022multi}, fraud detection~\cite{lu2022bright}, knowledge discovery~\cite{suarez2023templet,luzuriaga2021merging,rossi2022explaining,andreou2023using, jin2023graph,kannan2022exaflops,zheng2020dglke} and, etc.) are further solidifying the importance of GNNs.

Graph learning often follows an ego network-based computation graph (See Figure~\ref{fig:back}). 
Starting from a root node, the ego network of this node contains the full or a sampled subset of its in-neighbors. Moving to the next layer, we do the same for each selected in-neighbor. This process continues until we reach $k$-hop in-neighbors for a $k$-layer GNN. Since each node brings in multiple in-neighbors, {each ego network is a ``tree'' with the target node at the top and more and more nodes from the top to the bottom.}
This ``tree'' shape nature lets different ego networks share many nodes. 
Of note, P$^3$~\cite{gandhi2021p3}, Betty~\cite{yang2023betty}, FlexGraph~\cite{wang2021flexgraph} term this ego network as the GNN computation graph, multi-level bipartite graph, and hierarchical dependency graph, respectively.

\textbf{Problem definition. }
{A common task for GNNs is computing the embeddings of all nodes in an unseen graph. For example, fraud detection in e-commerce marketplaces views the millions of transactions in the past period as a graph~\cite{lu2022bright}.  In addition, particle physics experiments produce millions of high-dimensional measurements each second in the sensor array, which are represented as a graph because of their heterogeneity and sparsity in space~\cite{moreno2020jedi}.
}
We term this problem ``end-to-end inference for all nodes''. This daily update is required to reflect the latest changes in the graph. 
{Computing the embeddings entails applying a trained inductive GNN model to all graph nodes. }
Specifically, an end-to-end inference process involves (i) constructing the Compressed Sparse Row (CSR) from edge lists, (ii) partitioning the graph because the massive amount of activity records (edges formed) and active entities often result in graphs with billions of nodes and tens of billions of edges, and
(iii) computing the GNN embedding for each node.

While sharing opportunities arise during GNN computation because the GNN computation graph of each node is an ego network, we observe that this new ``end-to-end inference for all node'' maximizes such a sharing opportunity when compared with the inference of a subset of nodes (see DGI~\cite{yin2023dgi}) or training (HAG~\cite{jia2020redundancy} and P$^3$~\cite{gandhi2021p3}). For training, a GNN model can only be applied to a batch of nodes. Subsequently, the GNN model will be updated for the next batch, which limits the sharing in a batch.

Traditional GNN inference endeavors fail to fully extract the benefits of sharing for two reasons: (i) SALIENT++~\cite{kaler2023communication} caches the node features and reuses them across ego networks. However, the sharing is limited by the cache hit ratio. To increase the hit ratio, one needs to either increase the cache size or employ complicated caching policies, both of which introduce overwhelming system overheads~\cite{lin2020pagraph,yang2019aligraph, liu2023bgl}.
(ii) An alternative is to merge ego networks, automatically allowing nodes in the merged computation graph to enjoy the sharing benefits. However, this method has prohibitively high memory demands. DGI~\cite{yin2023dgi} and P$^3$~\cite{gandhi2021p3} thus only permits a subset of ego networks to work together.
Such a design can only exploit the sharing benefits within each subset, leaving cross-subset sharing wasted. 
Therefore, P$^3$ and DGI can only utilize 33\% and 60\% sharing in a 3-layer model.

This paper introduces {\name}, the first GNN inference system that is dedicated to end-to-end all-node inference on billion-edge graphs and is distributed to maximize the sharing benefits. Particularly, {\name} encompasses three contributions:

First, we unveil and exploit an untapped sharing opportunity during sampling, and maximize the benefits from sharing during subsequent GNN computation. During sampling, we take a fundamentally different approach, which completely eliminates the pointer-chasing problem faced by, to the best of our knowledge, all existing sampling approaches (i.e., ego network-centric sampling). Specifically, for $k$-layer GNN inference of all nodes, we sample 1-layer ego network $k$ times for each node. Subsequently, we collect the ego networks of the same layer across all nodes together to formulate a 1-hop graph. This offers us $k$ 1-hop ego networks for $k$ layers. During subsequent GNN computation, we feed the feature tensors of nodes and edges through these $k$ 1-hop graphs to arrive at the final embedding for all nodes (see Figure~\ref{fig:1hop}). This method automatically enjoys all sharing benefits during GNN computation.

Second, we employ a lightweight 1-D graph and feature collaborative partition to partition the graph and introduce memory-saving and communication-efficient distributed primitives for distributed end-to-end inference on the partitioned graph.
Specifically, we choose 1-D graph and feature collaborative partitioning for two reasons: 

(i)  1-hop graphs, along with the node features, are too big to fit in a single machine; (ii) end-to-end inference only contains one forward iteration, which cannot afford advanced node assignment algorithms whose overhead would outweigh the benefits~\cite{liu2023bgl, md2021distgnn}. 
Further, communication during GNN requires excessive memory to store the received data, especially for 1-D partitioning, where most of the edges are across partitions. Therefore,  we customize our distributed memory-efficient primitives, including GEneral Matrix Multiply (GEMM), SParse-dense Matrix Multiply (SPMM), and Sampled Dense-Dense Matrix Multiply (SDDMM). 
Compared to recent endeavors~\cite{kurt2023communication,tripathy2020reducing}, primitives in {\name} reduce memory usage during communication and retain communication efficiency.

Third, we introduce three system implementation optimizations specifically for end-to-end inference (Section~\ref{sec:comm}). First, we partition the sparse tensors into subgroups. The computation and communication are performed subgroup by subgroup to reduce the peak memory consumption. Second, we judiciously schedule the subgroup computation and communications to pipeline the computations and communications. We overlap the communication and computation in each subgroup and schedule the communication to reduce the bubbles. 
These optimizations can improve the SPMM and SDDMM performance beyond $3.50\times$.
Third, we fuse the first layer of GNN inference with graph construction to reduce feature tensor exchange overhead by avoiding redistributing the feature tensor based on the partition results. We accelerate graph pre-processing by up to 21.05$\times$ and reduce its wall-clock time ratio in end-to-end inference from 85\% to 29\% for various graph datasets.

\section{Background}\label{sec:back}

\subsection{GNN inference: an ego network centric computation}

 \begin{figure}[ht]
    \centering
    \includegraphics[page=1,width=\columnwidth]{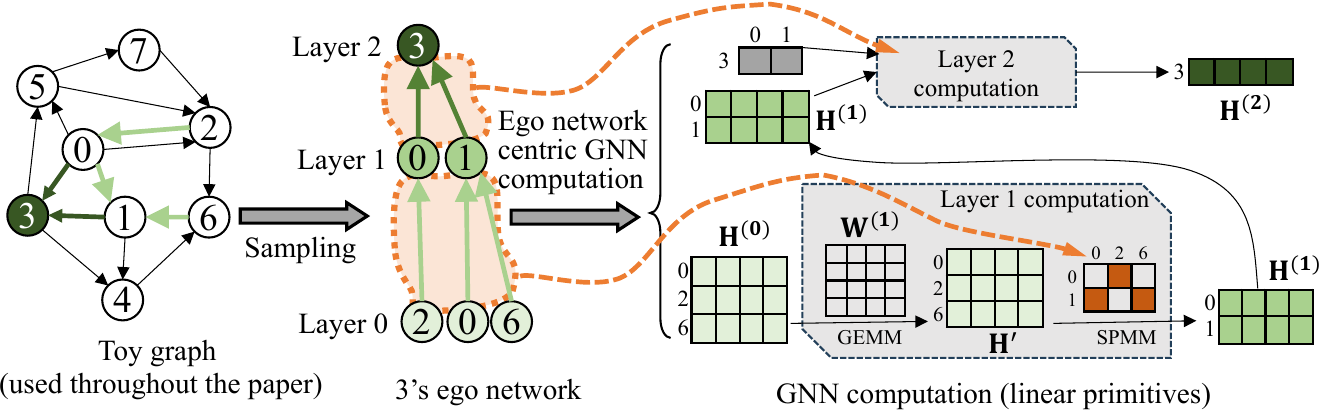}
    \caption{An example of inference computation workflow of {\name} for a 2-layer GCN.}
    \label{fig:back}
\end{figure}

\noindent
For a target node, GNN inference works on an ego network of this target node, which is extracted from the graph. The ego network consists of layers containing the neighbor nodes for one or more hops. The GNN computation progresses from layer 0 of the ego network towards this target node. One can rely on either sampling or the full graph to build this ego network.

\textbf{Ego network defined GNN computation. }
Node 3's ego network defines the subsequent GNN computation. While {\name} supports a variety of GNN models, for brevity, we use a well-known model, Graph Convolutional Networks (GCN)~\cite{kipf2016semi}, to explain this process. Usually, the computation is formalized as a series of linear primitives to take advantage of intra-ego network parallelism. The first step is deriving the ${\bf H}'$ through General Matrix-Matrix Multiplication (GEMM) from layers 0 to 1. Subsequently, we use Sparse-Dense Matrix Multiplication (SPMM) to aggregate features, arriving at ${\bf H}^1$. This process is repeated from layers 1 to 2 to calculate ${\bf H}^2$. ${\bf H}^2$ is the final embedding for target node 3.

\subsection{GNN partitioning methods}

\noindent
GNN projects mainly adopt three partitioning approaches during computation, i.e., 1-D and 2-D graph partitioning and feature partitioning. Of the three, 1-D and 2-D focus on partitioning the graph, while the last one distributes the feature tensor. 
As graphs and feature tensors continue to grow, partitioning is necessary for tackling large-scale GNNs. Particularly, feature tensors could grow significantly bigger than graphs. 
\textit{1-D partition} splits nodes based on node IDs and assigns a contiguous ID range to a partition. 
\textit{2-D partition} assigns each edge to one partition.
\textit{Feature partition} duplicates the graph and partitions the feature matrix into columns.

The choice of graph partitioning method significantly impacts the computation and communication of GNN. With 1-D partitioning, communication is required when accessing neighbors' features across machines during SPMM and SDDMM. In 2-D partitioning, each machine computes partial results of SPMM and communicates with machines holding the same row tiles. Feature partitioning turns GEMM into an outer product calculation, necessitating an expensive all-to-all reduction during GEMM, which can be the bottleneck for large-scale GNN training and inference.

\section{{\name} design \& implementation}
\label{sec:deal}

\subsection{Observations}

\begin{figure}[ht]
    \centering
    \includegraphics[width=.8\columnwidth]{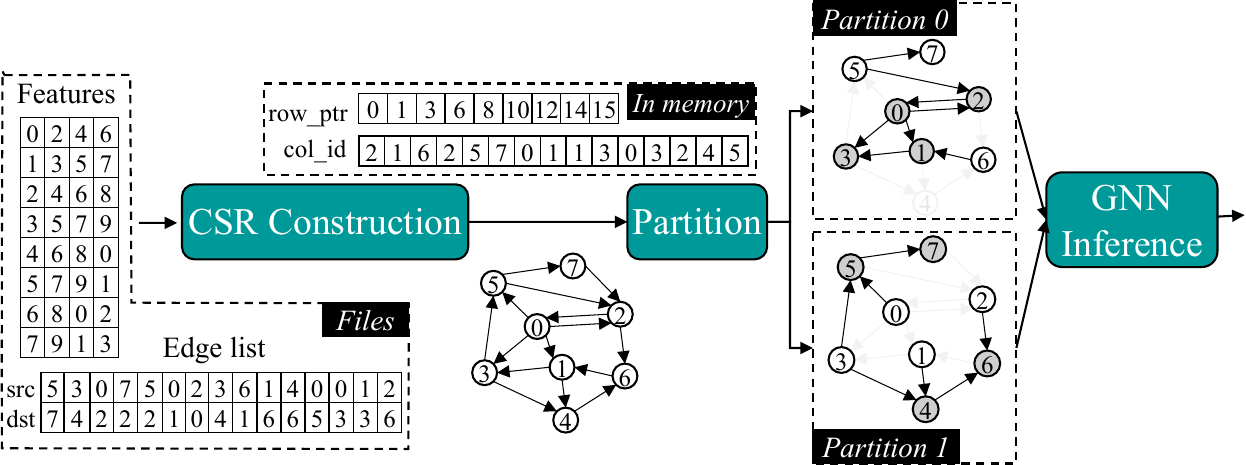}
    \caption{Illustration of end-to-end inference. }
    \label{fig:e2e}
\end{figure}

\noindent
Using the graph from Figure~\ref{fig:back}, Figure~\ref{fig:e2e} illustrates the four stages of end-to-end GNN inference.
First, the input graph is stored as an edge list on disk, and the graph generation entails reading the edge list and converting it to the graph data structure. Second, the partition algorithms are applied to the graph, dividing it into two partitions. The partitioned sub-graphs are stored in a shared file system accessible to all machines.
Third, every machine reads one graph partition in CSR format to memory. Lastly, machines perform the distributed GNN inference to compute the representation of all 8 nodes in the graph used for subsequent tasks.

\begin{figure}[ht]
% \vspace{-.2in}
\centering
\begin{tabular}{cc}  
    \subfloat[The time breakdown of end-to-end GNN inference, where we distribute the graph to four machines.]{
        \hspace{-.2in}
    \includegraphics[width=0.5\columnwidth]{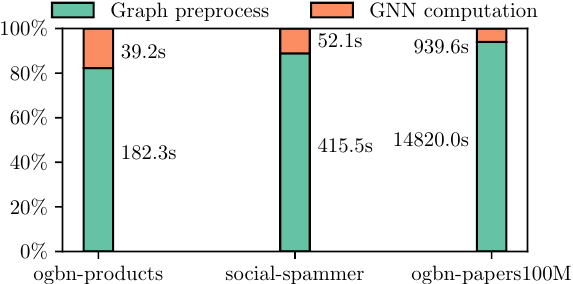}
    \label{fig:e2e_ratio}
    }
    &
    \subfloat[The peak memory consumption during inference with four partitions.]{
        \hspace{-.2in} 
    \includegraphics[width=0.5\columnwidth]{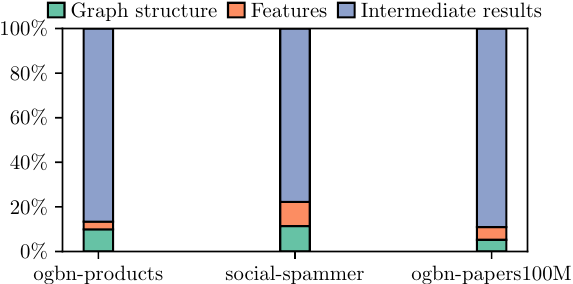}
    \label{fig:memory}
    }
\end{tabular}
\vspace{-.15in}
\caption{GNN partition analysis. 
\vspace{-.1in}}
\label{fig:motive_part}
\end{figure}

\textbf{Observation \#1. End-to-end inference requires a lightweight, co-designed topology-and-feature partitioning method.}
Figure~\ref{fig:e2e_ratio} presents the breakdown of the end-to-end inference time across three datasets, where the graph is generated and 1-D partitioned into four parts for inference. The results show that 86\% of the end-to-end time is spent on pre-processing, identifying it as a major efficiency bottleneck.
The reason is that GNN inference requires just a single epoch of forward computation to compute the representations for all nodes. This is different from training that performs forward and backward computation multiple times to obtain a converged model. 
Therefore, advanced ``time-consuming'' graph pre-processing algorithms (such as partitioning and reordering) might not be a good fit because the time saved during inference by advanced pre-processing steps (e.g., advanced partitioning like METIS~\cite{karypis1997parmetis}) could be shorter than the time spent in pre-processing. 

Neither graph partition nor feature partition alone would meet the requirements. 
(i) On the one hand, graph partition alone would incur excessive memory consumption. Specifically, during ego network computation, when the number of layers in the ego network grows, most nodes become cross-partition nodes. One would need a space to hold these features as the intermediate results for computation. 
Figure~\ref{fig:memory} shows the memory consumption of GNN inference with four partitions. For example, when the ogbn-products~\cite{Bhatia16} graph is partitioned into four parts, one partition receives messages from 80\% of the total nodes in the graph, leading to more than 380 GB memory footprint on one machine.
(ii) On the other hand, feature partition alone will introduce excessive all-to-all communications during GEMM and SDDMM computation, which are illustrated in Figures~\ref{fig:GEMM} and~\ref{fig:SDDMM}.

\textit{{\name} design}. 
We introduce node feature partitioning, in addition to 1-D graph partition, to mitigate excessive memory consumption and communication costs (Section~\ref{subsec:part}). First off, this design is lightweight. Second, it will divide a big primitive into several smaller ones, like one GEMM on big matrices, into GEMM on several smaller ones (Section~\ref{sec:primitives}). This strategy not only bounds the communication required for each group compared to the entire set of primitives but also significantly lowers peak memory usage.

 \begin{figure*}[ht]
    \centering
    \includegraphics[page=1,width=.9\textwidth]{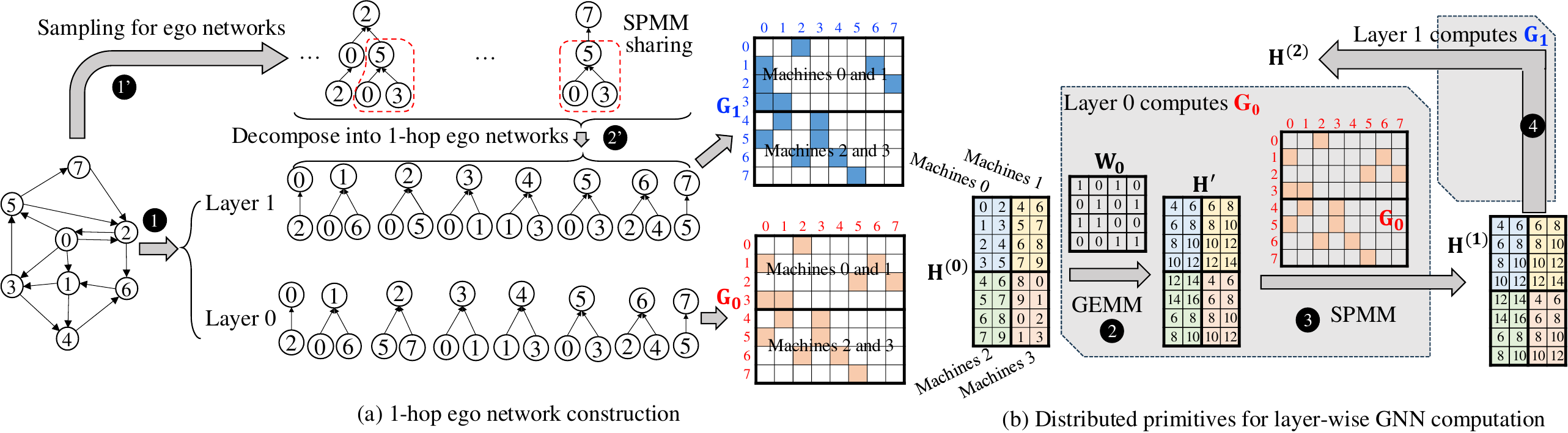}
    \vspace{-.1in}
    \caption{{\name} workflow with layer-by-layer inference design.
    % \vspace{-.2in}
    }
    \label{fig:1hop}
\end{figure*}

 \begin{figure}[H]
    \centering
    \includegraphics[width=.8\columnwidth]{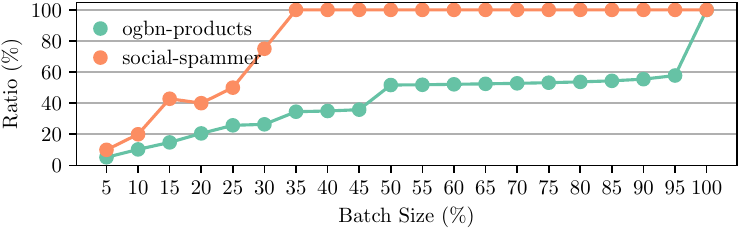}
    \caption{The leveraged sharing opportunity in different inference batch sizes (percentage of all nodes).
    }
    \label{fig:share}
\end{figure}

\textbf{Observation \#2. All-node inference presents new sharing opportunities and challenges.} 
While computing the GNN inferences for different target nodes together, offering sharing opportunities is not new (see, SALIENT++~\cite{kaler2022accelerating}, DGI~\cite{yin2023dgi} and P$^3$~\cite{gandhi2021p3}), the amount of sharing for all-node inferences is unprecedented, which renders new research opportunities and challenges.

The key insight is that only batching enough inferences will offer considerable sharing opportunities, which also means significant memory space consumption.  
Figure~\ref{fig:share} illustrates the sharing opportunities for a 3-layer GNN while increasing the batch sizes for sparse (obgn-products) and dense (social-spammer) graphs. Particularly for sparse graphs like ogbn-products, full sharing is only achieved with a single batch due to low connectivity, which causes partial batches to include nodes across batches redundantly. For dense graphs like social-spammer, high connectivity allows each batch to cover a distinct part of the graph. While seems promising, it actually indicates increasing batch size will consume enormous memory space. For instance, a machine with 256 GB memory can only accommodate the batch size up to 6\% of the nodes (149K) for ogbn-products, 0.12\% (103K) for social-spammer.

\textit{{\name} design}. 
We propose processing all-node inference in a single batch to extract the sharing benefits fully. First, we implement GNN operations as distributed linear algebra primitives and further optimize them to be communication efficient (Section~\ref{sec:primitives}). Second, we strategically partition the GNN primitives to reduce the extra communication from group-by-group execution.
Additionally, we implement a pipelining strategy for communicating these groups, which helps in overlapping communication and computation. This reduces the waiting time for data transfer and ensures better resource utilization (Section~\ref{sec:comm}).

\vspace{-.1in}
\subsection{{\name} Workflow Overview}

A natural way of taking advantage of the sharing in observation \#1 would take \circled{1'} - \circled{2'} option in Figure~\ref{fig:1hop}. That is, we first obtain all the multi-hop ego networks. Subsequently, we break them into 1-hop ego networks. Finally, we remove the duplicated 1-hop ego networks. For instance, the 1-hop ego network of 5 is shared across the ego networks for nodes 2 and 7. Therefore, we only store that 1-hop ego network once. 

{\name} goes further by completely avoiding building the multi-hop ego networks. In fact, we compute the embeddings for all target nodes without recovering the multi-hop ego network. 
Of note, {\name} can also work for complete graph-based embedding updates (i.e., each one-hop ego network contains the entire neighborhood). 

As shown in \circled{1} of Figure~\ref{fig:1hop}, we directly sample 1-hop ego networks for all nodes. For each layer, we will collectively store all of these 1-hop ego networks as a graph. For instance, the layer 0 graph is stored as $\bf G_0$. Similarly, layer 1 as $\bf G_1$. 
We sample two 1-hop ego networks for every node for two GNN layers, as shown in Figure~\ref{fig:1hop}(a). 
The 8-node graph with a 2-layer GNN leads to 16 ego networks, which can be combined to form the multi-hop ego network.
For example, the 2-hop ego network of node 2 comprises the 1-hop ego network of node 2 at layer 2, nodes 0 and 5 at layer 1. 
In Figure~\ref{fig:1hop}, the aggregation of 1-hop ego network in every row is unique, while the sampling in each column accesses the neighbors of the same node. Meanwhile, in each row, ego networks may share the neighbors. Therefore, we can maximize the sharing within ego networks by sampling column-wise and the sharing between ego networks by computing row-wise.
Of note, if we work on the complete graph, we will use the complete graph $\bf G$ as $\bf G_0$ and $\bf G_1$ for the subsequent computations.

We sample the 1-hop ego network in one column together to leverage the sharing in sampling.
In particular, sampling from a distribution requires a data structure representing the distribution. Building and accessing this data structure leads to the major overhead of sampling. For example, when sampling multiple neighbors without replacement, a tree is built where each branch indicates the sampling space after the certain neighbor is picked, and the sampling is to traverse the tree randomly. 
Therefore, when sampling the same target node for different GNN layers, the same data structure can be reused, saving construction costs.
The sampled 1-hop ego networks are stored as an edge list for the computation.

Computation sharing is achieved in two ways: First, the node projection GEMM of the same node is shared. For instance, in layer 1, neighbor 0 of 1, 2, 3, and 5 are shared (\circled{2}). 
Second, the aggregation SPMM of the same node is shared. As shown in step \circled{3}, node 5's aggregation is shared for target nodes 2 and 7. 
We notice that certain nodes might not appear in any multi-hop ego networks due to the neighbor sample, but we still sample and compute its 1-hop network to simplify the implementation. We notice that nearly all nodes will appear on each layer of the ego networks because the number of dependency nodes increases exponentially at each layer.

\subsection{Topology and feature co-designed partition}
\label{subsec:part}

\begin{figure}[H]
    \centering
    \includegraphics[width=1\columnwidth]{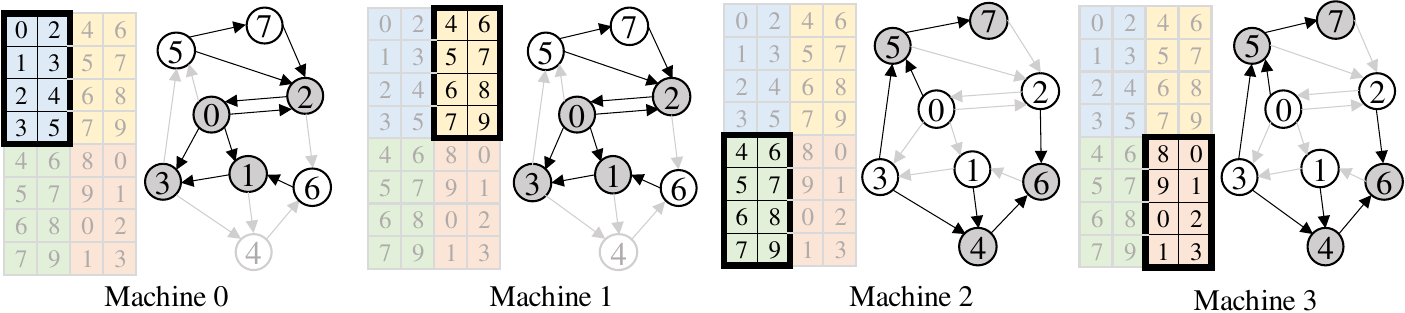}
    \caption{{\name} partitioning strategy.}
    \label{fig:partition}
    % \vspace{-.1in}
\end{figure}

\noindent
{\name} partitions both the graph topology and node features to curb the memory consumption and the time spent on partitioning. Specifically, we adopt 1-D graph partition such that each machine obtains all the in-neighbors of a disjoint equal range of nodes. Further, we distribute the features of each partition across multiple machines. 
Figure~\ref{fig:partition} explains how the same toy example would be distributed across four machines. Specifically, we partition nodes range 0 - 3 and 4 - 7 into two partitions. Machines 0 and 1 both host one copy of the edge list of Partition 0. Machines 2 and 3 both host one copy of the edge list of Partition 1.
In the meantime, we partition the features of each node between machines hosting the same partition. Therefore, machine 0 
will be responsible for the first two feature dimensions of nodes 0 - 3, and machine 1 for the second two dimensions of nodes 0 - 3, similar to the other two machines.

\begin{figure*}[ht]
    \centering
    \includegraphics[page=1,width=0.95\textwidth]{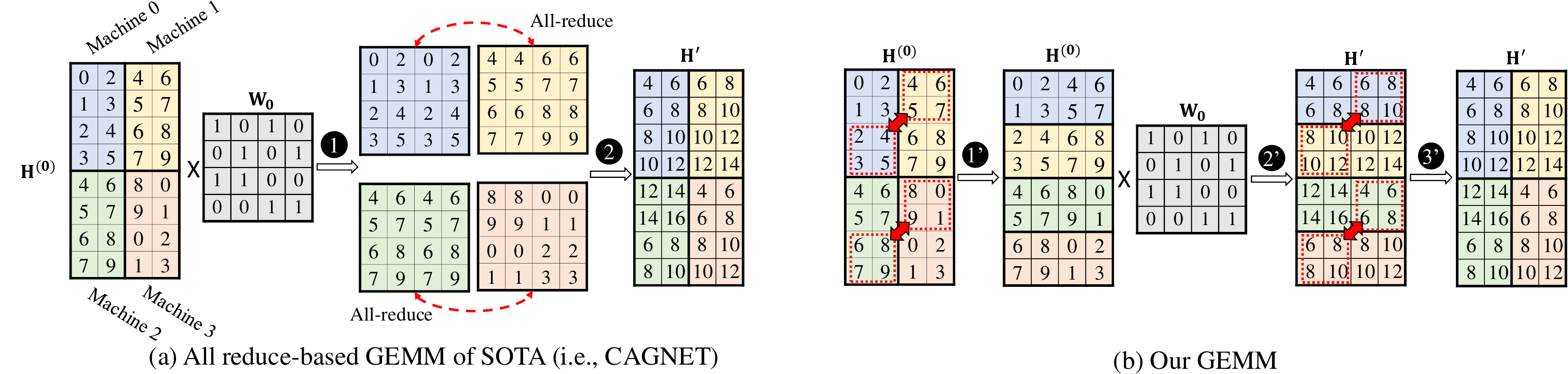}
    \vspace{-.1in}
    \caption{An example of distributed GEMM with four machines. (a) SOTA GEMM. (b) Our GEMM.
   \vspace{-.1in} 
    }
    \label{fig:GEMM}
\end{figure*}

Our partitioning approach is more lightweight and communication efficient than both traditional 2-D-based graph partition and feature partition: 
(i) 2-D partition, splits the adjacency matrix into tiles in both row and column directions. Therefore, during the SPMM primitive, each machine computes the partial results and needs to communicate to other machines with the tiles in the same row. 
Our partitioning stores the full rows on a machine to avoid such distributed aggregation requirements.
(ii) Feature partition, distributes the features of nodes across machines so that each machine stores the entire column of the features. As a result, GEMM computation becomes an outer product calculation. This will result in an all-to-all reduction during GEMM computation, which could be very expensive. 
In our approach, only the machines with the same rows of feature tensors need to communicate, reducing the total communication size.
More details are presented in Section~\ref{sec:primitives}.

\subsection{{\name} distributed GNN primitives} \label{sec:primitives}

\textbf{GEMM} multiplies the partitioned feature matrix $\bf H^{(0)}$ with the weight matrix $\bf W_0$. We let each machine own a part of the feature matrix and the full weight matrix because $\bf W_0$ is significantly smaller than $\bf H^{(0)}$ in GNN.

\textit{SOTA GEMM}.
Figure~\ref{fig:GEMM}(a) illustrates the design in existing SOTA all reduce-based GEMM, i.e., CAGNET~\cite{tripathy2020reducing}. 
At step \circled{1}, every machine multiplies its local tile with the associated rows in the weight matrix. In the example, machine 0 multiplies with the first 2 rows, and machine 1 multiplies with the second 2 rows. Each machine derives a $4\times 4$ matrix.
Subsequently, at step \circled{2}, machines sharing the rows aggregate the columns from each other to compute the resultant columns.

CAGNET's GEMM faces two drawbacks: excessive communication cost and memory consumption. 
(i) In Figure~\ref{fig:GEMM}(a), each machine has to receive partial results from all the other machines for the tile this machine is responsible for. 
(ii) For space consumption, CAGNET creates the intermediate result of size $4\times 4$ on each machine before aggregation.
Assuming $\bf H^{(0)}$ has $N$ rows and $D$ columns, and $H^{(0)}$ is partitioned into $P\times M$ partitions, that means we have $PM$ machines. Therefore, each machine works on $\frac{ND}{PM}$ entries. During GEMM, each machine receives $\frac{ND}{PM}(M-1)$ in entries for the feature tensor, and the memory footprint increases from $\frac{ND}{M}$ to $\frac{N D}{P}$.

\textit{Our GEMM}.
Figure~\ref{fig:GEMM}(b) introduces our design, significantly reducing the memory costs and the communication overhead CAGNET faces. The key idea is to avoid creating large intermediate results on each machine,  as well as fully leverage the benefits of the duplicated $\bf W_0$ matrix. In the example, at step \circled{1'}, machine 0 partitions its $4\times 2$ tile of $\bf H^{(0)}$ into two $2\times 2$ tiles. Then, it keeps the first tile and sends/receives the remaining tiles to/from the other machines. After that, machine 0 owns a $2\time 4$ tile for the first two rows and uses that to multiply with $\bf W_0$ to arrive at the first two rows of \textbf{H'} (\circled{2'}). The final step (\circled{3'}) performs the same communication pattern as the first step so that machine 0 could, again, maintain the $4\times 2$ tile of the feature matrix (\textbf{H'}).

We implement a ring-based all-to-all communication to pipeline the computation. Using step \circled{1'} as an example, for the first entry, four machines form a logical ring: machines $0 \rightarrow 1 \rightarrow ... \rightarrow (M-1) \rightarrow 0$. This process continues until we arrive at the row-wise distributed ${\bf H^{(0)}}$. In this example, we only have 2 machines per sharing each row of ${\bf H^{(0)}}$. So it is simply a Ping-Pong exchange.  The communication of step \circled{3'} is similar.
Since we break the all-to-all communication into $M-1$ stages, we can overlap the communication with the computation. For example, machine 0 multiplies $\bigl[$\colorbox[HTML]{9BC4E2}{$\begin{smallmatrix}0 & 2 \\ 1 & 3\end{smallmatrix}$}$\bigr]$ with $\bf W_0$ while receiving $\bigl[$ \colorbox[HTML]{FFE5B4}{$\begin{smallmatrix}4 & 6 \\ 5 & 7\end{smallmatrix}$}$\bigr]$ from machine 1. This will further reduce the size of the intermediate result.

\begin{table}[ht]
  \renewcommand{\arraystretch}{1.3}
  \caption{Memory and communication costs of GEMM.}
  % \resizebox{0.7\columnwidth}{!}{
  {\scriptsize
  \begin{tabular}{ccc}
    \toprule
     Method & Memory & Communication\\
    \midrule
    SOTA & $\frac{ND}{P}$& $\frac{ND}{PM}(M-1)$ \\
    Ours& $\frac{N D}{PM^2}$ & $2\frac{ND}{PM^2}(M-1)$\\
    \bottomrule
  \end{tabular}
  }
    \label{tab:gemm}
    \vspace{-.1in}
\end{table}

We reduce memory by $M^2\times$ and communication costs by $\frac{M}{2}\times$ compared with SOTA as shown in Table~\ref{tab:gemm}.
At step \circled{1'}, one machine splits its partition into $M$ blocks with the size of each block as $\frac{ND}{PM^2}$. Each machine will send $M-1$ blocks to the $M-1$ rest of machines. Therefore, the communication size of one machine is $2\frac{ND}{PM^2}(M-1)$ because it happens in \circled{1'} and \circled{3'}.

\begin{figure}[b]
\vspace{-.2in}
    \centering
    \includegraphics[page=1,width=1\columnwidth]{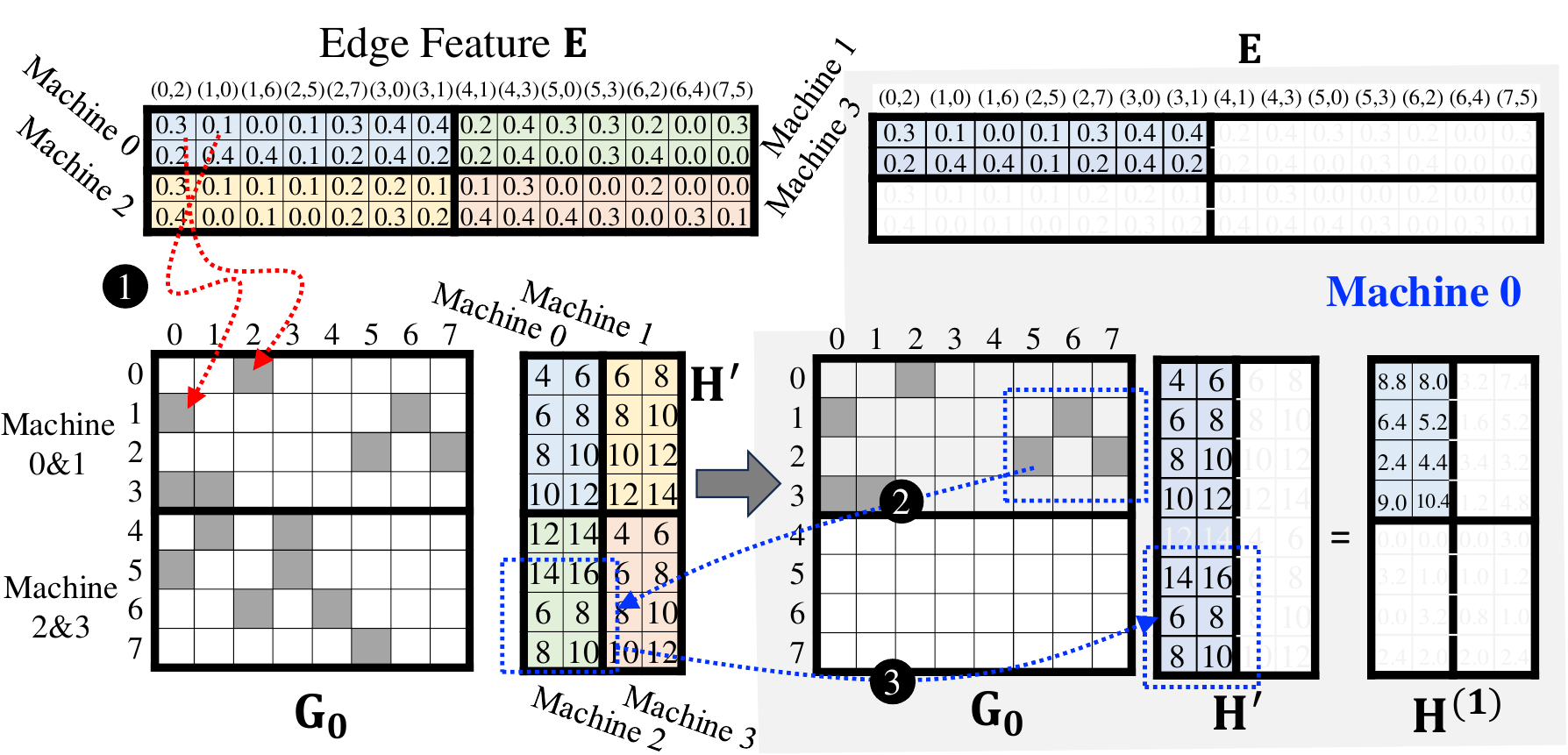}
    \caption{Our distributed three-tensor SPMM under 1-D partitioning strategy for ${\bf G_0}$ on machine 0.
    \vspace{-.1in}
    }
    \label{fig:SPMM}
\end{figure}

\textbf{SPMM} multiplies the node embedding matrix $\bf H^{'}$ with the edge features $\bf E$ based on the graph connectivity ${\bf G_0}$. 
Formally, $\bf H^{(l)}[][{\text{i}}]= multiply_G(E[\text{i}][], \bf H'[][\text{i}])$. 
Figure~\ref{fig:SPMM} uses ${\bf G_0}$ as an example, it multiplies the $i$-th row of \textbf{E} with $i$-th column of $\bf H'$ following Sparse-Matrix Vector Multiplication (SpMV) fashion. During multiplication, \textbf{E} is shaped into ${\bf G_0}$ for proper matrix multiplication. As shown in \circled{1}, the first row of $\bf E$, i.e., \{0.3, 0.1, ..., 0.3\} is loaded into corresponding edge locations of ${\bf G_0}$ to multiply with $\bf H'$.

\textit{Our SPMM}. {\name}'s SPMM communicates the $\bf H'$ matrix to realize distributed SPMM. As shown in Figure~\ref{fig:SPMM}, machines 0 and 1 hold the top half of ${\bf G_0}$ while 2 and 3 are the bottom. The $\bf H'$ matrix follows the partitioning in GEMM. Each machine holds the edge features of the edges (non-zeros) within its ${\bf G_0}$ part, aligning with the feature partition of $\bf H'$. Using machine 0 as an example, it is responsible for computing the blue tile of $\bf H'$. As shown in the blue dotted box and dashed arrows, during SPMM, machine 0 sends the non-zeros column IDs (5,6,7) to machine 2 (\circled{2}), and machine 2 returns rows 5-7 of $\bf H'$ (\circled{3}), i.e., $\bigl[$\colorbox[HTML]{ACE1AF}{$\begin{smallmatrix}14 & 16 \\ 6 & 8\\8&10\end{smallmatrix}$}$\bigr]$. After that, machine 0 computes the resultant tile $\bf H^{(1)}$ with its local ${\bf G_0}$, $\bf E$, and $\bf H'$.

\textit{Exchange $\bf G_0$}.
An alternative approach to fulfilling the communication is to exchange the sparse graph ${\bf G_0}$. Using machine 0 as an example, it first multiplies columns 0-3 of ${\bf G_0}$ with its ${\bf H'}$ tile as a partial result. It then sends its ${\bf G_0}$ tile and the associated edge features to machine 1. Machine 1 performs multiplication with its local ${\bf H'}$ tile and returns the resultant partial product to machine 0 to aggregate as the final $\bf H^{(1)}$. Although this method reduces the initial communication volume by transmitting only the graph structure and edge features, the second communication phase involves transferring partial results, whose size is comparable to that of the $\bf H'$ tile. Consequently, the overall communication cost exceeds that of our SPMM approach.

\begin{figure}[h]
    \centering
    \includegraphics[page=1,width=.95\columnwidth]{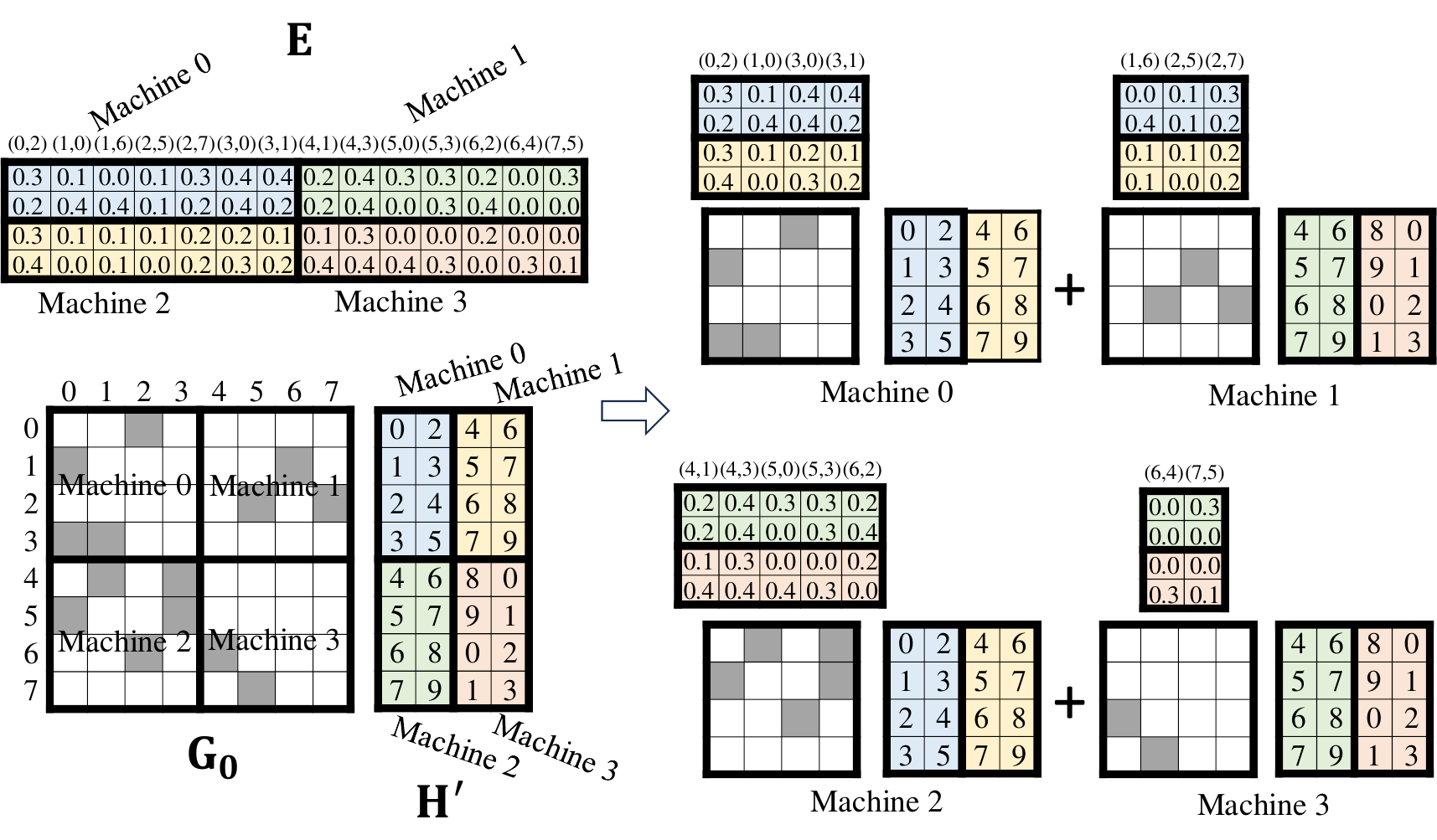}
    \caption{SOTA 2D-based SPMM.
    \vspace{-.2in}
    }
    \label{fig:dist}
\end{figure}

\textit{SOTA 2D-based SPMM}~\cite{tripathy2020reducing,peng2022sancus,selvitopi2021distributed,kurt2023communication}.
As depicted in Figure~\ref{fig:dist}, the sparse matrix (${\bf G_0}$) and the feature matrix ($\bf H{'}$ and $\bf E$) are partitioned in 2D and distributed across four machines. During SPMM, machine 0 first receives the $\bf H{'}$ tile from machine 1. Machine 0 then performs local SPMM with its local edge features $\bf E$ to compute partial results of $\bf H^{(l)}$. After that, machine 0 receives the partial result of columns 0 and 1 from machine 1 to derive the final results. Both {\name} and 2D-based SPMM receive a $4\times 2$ tile of $\bf H{'}$ from machine 3 and machine 1, respectively. Since both approaches can send the non-zero index first to reduce the actual transferred features, they initially have similar communication costs. However, 2D-based SPMM needs to send its partial $4\times 2$ results for columns 2 and 3 to machine 1, which is not required in {\name}'s approach.

\begin{table}[ht]
  \renewcommand{\arraystretch}{1.6}
  \caption{Memory and communication costs of SPMM. (Note memory consumption is the same as communication cost)}
  {\scriptsize
  \begin{tabular}{ccc}
    \toprule
    Method & Memory & Communication\\
    \midrule
    Ours & - & $\frac{ZN(P-1)}{P^2}+\frac{N(P-1)}{P^2}\frac{D}{M}$ \\
    Exchange ${\bf G_0}$ & - & $\frac{ZN(P-1)}{P^2}\frac{D}{M} + \frac{ND}{PM}$ \\
    2D-based SPMM  & - & $\frac{N(P-1)}{P^2}\frac{D}{M} + \frac{ND(M-1)}{PM}$\\
    
    \bottomrule
  \end{tabular}
  }
    \label{tab:spmm}
    \vspace{-.1in}
\end{table}

The communication size is determined by two messages. Consider the distributed SPMM multiplying an $N\times N$ $\bf G_0$ with an $N\times D$ $\bf H'$, which has $P$ parts for rows and $M$ parts for columns (same as $\bf H^{(0)}$ in GEMM). Assuming that each column has $Z$ non-zeros on average, every machine receives $\frac{ZN(P-1)}{P^2}$ non-zeros from other machines (\circled{2}), which contains $\frac{N(P-1)}{P^2}$ unique columns. Further, since each machine receives the $\frac{D}{M}$ features for every non-zero column in $\bf H'$, the communication size is $\frac{N(P-1)}{P^2}\frac{D}{M}$ (\circled{3}). Similarly, for exchanging graphs, the $\bf G_0$ leads to $\frac{ZN(P-1)}{P^2}\frac{D}{M}$ communication and the partial result leads to $\frac{ND(M-1)}{PM}$ communication. 
For 2D-based SPMM, the extra aggregation leads to $\frac{ND(M-1)}{PM}$ communication. Compared with exchanging $\bf G_0$, both our first term (graph) and the second term (features) are smaller. Further, compared with 2D-based SPMM, the second term of 2D-based SPMM is much larger than ours. Together, our design is more communication efficient.

\begin{figure}[ht]
    \centering
    \includegraphics[page=1,width=0.95\columnwidth]{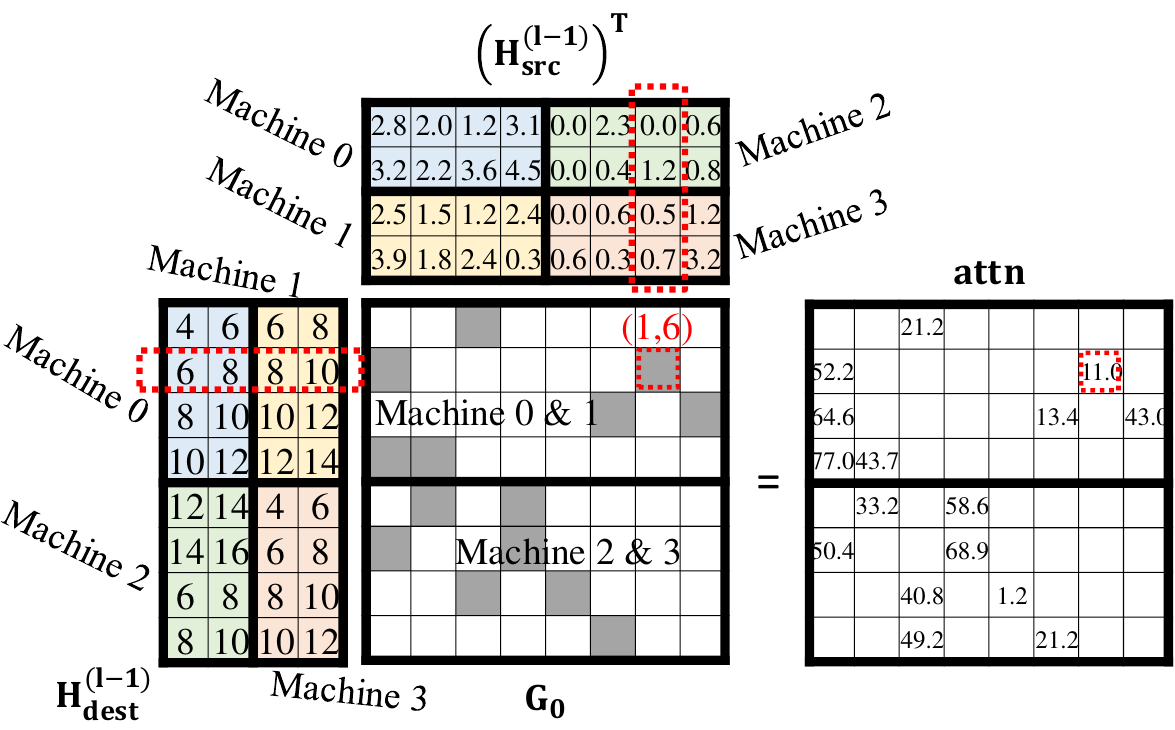}
    \caption{Our distributed SDDMM on ${\bf G_0}$. 
    % \hang{\textbf{G} not included in Fig}
    \vspace{-.15in}
    }
    \label{fig:SDDMM}
\end{figure}

\textbf{SDDMM} primitive uses the source features matrix $\bf H^{(l-1)}_{src}$ and the destination feature matrix $\bf {H^{(l-1)}_{dest}}$ to derive the edge attention based on the adjacency matrix ${\bf G_0}$. Formally, $\textbf{attn}={\bf G_0}\odot(\bf {H^{(l-1)}_{dest}}\cdot   ({\bf H}^{(l-1)}_{src})^T  )$.
As shown in Figure~\ref{fig:SDDMM}, the highlighted nonzero $(1, 6)$ computation in ${\bf G_0}$ is the dot-product between row 1 in $\bf {H^{(l-1)}_{dest}}$ and column 6 in $\bf (H^{(l-1)}_{src})^T$. 
Practically, only the positions with non-zeros associated in ${\bf G_0}$ are computed, and the result sparsity is identical to ${\bf G_0}$.
In distributed SDDMM, the computation of any nonzero entries would involve feature matrices from multiple machines.
For instance, computing entry $(1, 6)$ requires data from four machines (highlighted in red dashed boxes in $\bf{H^{(l-1)}_{dest}}$ and $\bf (H^{(l-1)}_{src})^T$).

We propose an output-oriented task scheduling with corresponding communication patterns. Specifically, we assign machines storing the non-zeros in ${\bf G_0}$ to compute the corresponding $\bf attn$ results. This strategy ensures that the results for each non-zero element are co-located with the sparse matrix after the SDDMM operation. When multiple machines store the same portion of the sparse matrix, we consider two approaches: (i) duplicating the computation across machines or (ii) distributing the computation of non-zeros among machines and subsequently exchanging the results. 

\textit{Approach (i). }
Using machine 0 as an example, approach (i) requires all four features of nodes 5-7 in $\bf (H^{(l-1)}_{src})^T$ from machines 2 and 3, which is 12 values
$\bigl[ \begin{smallmatrix}\colorbox[HTML]{ACE1AF}{\footnotesize 2.3\ \  0.0\ \  0.6}\\\colorbox[HTML]{ACE1AF}{\footnotesize 0.4\ \ 1.2\ \ 0.8}\\\colorbox[HTML]{FFE4E1}{\footnotesize 0.6\ \ 0.5\ \ 1.2}\\\colorbox[HTML]{FFE4E1}{\footnotesize 0.3\ \ 0.7\ \ 3.2}\end{smallmatrix}\bigr]$. We also need all 8 values of $\bf H^{(l-1)}_{dest}$, i.e., $\bigl[$\colorbox[HTML]{FFE5B4}{$\begin{smallmatrix}6 & 8\\8 & 10\\10 & 12\\12 & 14\end{smallmatrix}$}$\bigr]$ from machine 1, and 6 values of $\bf (H^{(l-1)}_{src})^T$, i.e., $\bigl[$\colorbox[HTML]{FFE5B4}{$\begin{smallmatrix}2.5 & 1.5 & 1.2\\3.9 & 1.8 & 2.4\end{smallmatrix}$}$\bigr]$ from machine 1.
The communication size is 26.

\textit{Approach (ii)}, we let machine 0 compute the non-zeros in $\bf attn$ rows 0-1, and machine 1 in rows 2-3. Therefore, machine 0 receives $\bigl[ \begin{smallmatrix}\colorbox[HTML]{ACE1AF}{\footnotesize 0.0}\\\colorbox[HTML]{ACE1AF}{\footnotesize 1.2}\\\colorbox[HTML]{FFE4E1}{\footnotesize 0.5}\\\colorbox[HTML]{FFE4E1}{\footnotesize 0.7}\end{smallmatrix}\bigr]$, $\bigl[$\colorbox[HTML]{FFE5B4}{$\begin{smallmatrix}2.5 \\3.9\end{smallmatrix}$}$\bigr]$, and $\bigl[$\colorbox[HTML]{FFE5B4}{$\begin{smallmatrix}1.2\\ 2.4\end{smallmatrix}$}$\bigr]$ from $\bf (H^{(l-1)}_{src})^T$, and $\bigl[$\colorbox[HTML]{FFE5B4}{$\begin{smallmatrix}6 & 8\\8 & 10\end{smallmatrix}$}$\bigr]$ in $\bf {H^{(l-1)}_{dest}}$. 
After the computation, machine 0 receives rows 2 - 3 in $\textbf{attn}$ from machine 1. In total, machine 0 receives 17 values.
In the meantime, machine 1 receives
$\bigl[ \begin{smallmatrix}\colorbox[HTML]{ACE1AF}{\footnotesize 2.3}\\\colorbox[HTML]{ACE1AF}{\footnotesize 0.4}\\\colorbox[HTML]{FFE4E1}{\footnotesize 0.6}\\\colorbox[HTML]{FFE4E1}{\footnotesize 0.3}\end{smallmatrix}\bigr]$, $\bigl[ \begin{smallmatrix}\colorbox[HTML]{ACE1AF}{\footnotesize 0.6}\\\colorbox[HTML]{ACE1AF}{\footnotesize 0.8}\\\colorbox[HTML]{FFE4E1}{\footnotesize 1.2}\\\colorbox[HTML]{FFE4E1}{\footnotesize 3.2}\end{smallmatrix}\bigr]$, and $\bigl[$\colorbox[HTML]{9BC4E2}{$\begin{smallmatrix}2.8 & 2.0\\3.2 & 2.2\end{smallmatrix}$}$\bigr]$ in $ ({\bf H}^{(l-1)}_{src})^T$, and $\bigl[$\colorbox[HTML]{9BC4E2}{$\begin{smallmatrix}8 & 10\\10 & 12\end{smallmatrix}$}$\bigr]$ in ${\bf H}^{(l-1)}_{dest}$. 
For results, it receives rows 0 - 1 from machine 0. The total \# of received values is 19. Further, the two machines communicate in parallel.
Therefore, approach (ii) leads to fewer communications in this example.

\begin{table}[ht]
  \renewcommand{\arraystretch}{1.3}
  \caption{Memory and communication costs of SDDMM. (Note memory consumption is the same as communication cost)}
  {\scriptsize
  \begin{tabular}{ccc}
    \toprule
     Method & Memory & Communication\\
    \midrule
    Approach (i) & -  &  $(M+MP - 2)\frac{ND}{MP}$\\
    Approach (ii)  & - &  $(M+MP - 2)\frac{ND}{M^2P}$ + $\frac{NZ(M-1)}{PM}
    $ \\
    \bottomrule
  \end{tabular}
  }
    \label{tab:sddmm}
    % \vspace{-.2in}
\end{table}

We choose approach (ii) for reduced communication size. 
Similar to SPMM assumptions, we assume $\bf H^{(l-1)}_{dest}$ and $\bf H^{(l-1)}_{src}$ are $N\times D$ dense with $P$ partitions in rows and $M$ partitions in columns, and $\bf{G_0}$ is $N \times N$ with $Z$ non-zeros per column on average. So \# of machines = $MP$. For approach (i), each machine needs to access $M-1$ and $MP -1$ machines from $\bf H^{(l-1)}_{dest}$ and $\bf H^{(l-1)}_{src}$, respectively, so the total communication is $(M+MP - 2)\frac{ND}{MP}$. 
For approach (ii),
each machine computes $\frac{N}{MP}$ rows instead of $\frac{N}{P}$ rows in approach (i). Therefore, the total communication is reduced to $(M+MP - 2)\frac{ND}{M^2P}$. Further, approach (ii) requires communicating the $\frac{M-1}{M}$ ratio of the nonzeros in $\bf attn$, which leads to $\frac{NZ(M-1)}{PM}$ communications. In total, approach (ii) performs $(M+MP - 2)\frac{ND}{M^2P}$ + $\frac{NZ(M-1)}{PM}$ communications. When $M$ increase, the communication size of the input in Approach (ii) is reduced faster than that of Approach (i), which supports our choice of Approach (ii).

\subsection{{\name} system optimizations} \label{sec:comm}

\begin{figure}[h]
  \centering
  \includegraphics[width=0.6\columnwidth]%
    {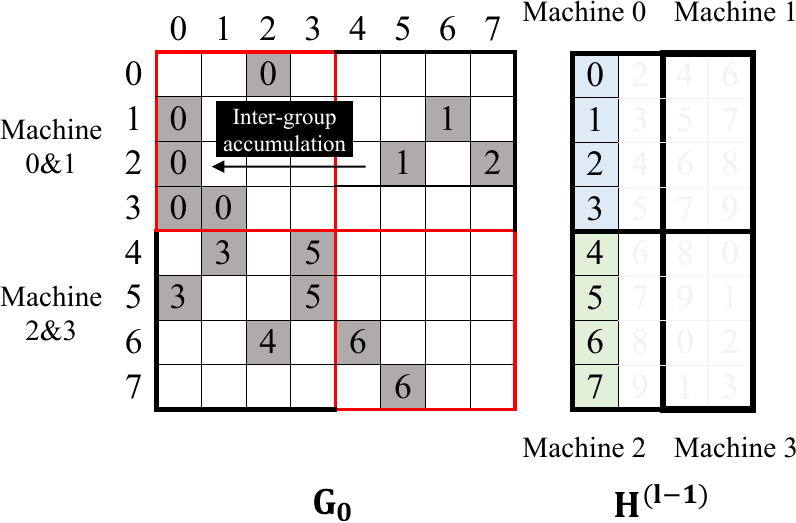}% picture filename
      \caption{Partitioned SPMM computation. The number in ${\bf G_0}$ represents which group this nonzero belongs to. ``0'', for examples, means this entry belonging to group 0. }
      \label{fig:nz}
\end{figure}

\noindent
\textbf{Partitioned communication.} 
Figure~\ref{fig:nz} exemplifies our partitioning strategy, which consists of two steps. First, we assign the non-zeros from ${\bf G_0}$ that multiply with local node features of $\bf H^{(l-1)}$ into one group. For example, the non-zeros in the top-left tile in ${\bf G_0}$ are local for machine 0 and machine1.
Second, we partition the other non-zeros based on their column IDs. 
{In particular, we sort the column ID array in CSR and assign non-zeros in adjacent columns into groups because they fall within a small range in the sorted column ID array. }
Using machine 0 as an example, we put the non-zeros in columns 5 and 6 of ${\bf G_0}$ into group 1 and the remaining into group 2. 
As a result, the computation of ${\bf G_0}$ is partitioned into 6 groups, where machines 0 \& 1 compute groups 0-2, and machines 2 \& 3 compute groups 3-6.
In each group, we communicate the needed features and multiply them with the edge features of the non-zeros in the group.

When non-zeros of the same row in ${\bf G_0}$ are partitioned into different groups, we cache the results of each row and perform inter-group accumulation. In the example, we cache the partial results of rows 0-3 derived by the group 0. Then, for group 1, we accumulate the results of $(4,1)$ and $(4,3)$ in ${\bf G_0}$ to the cache of rows 1 and 3, respectively. The SDDMM computation is similar. We focus on one group of non-zeros at a time and perform the computations.

\begin{figure}[ht]
    \centering
    \includegraphics[page=1,width=.7\columnwidth]{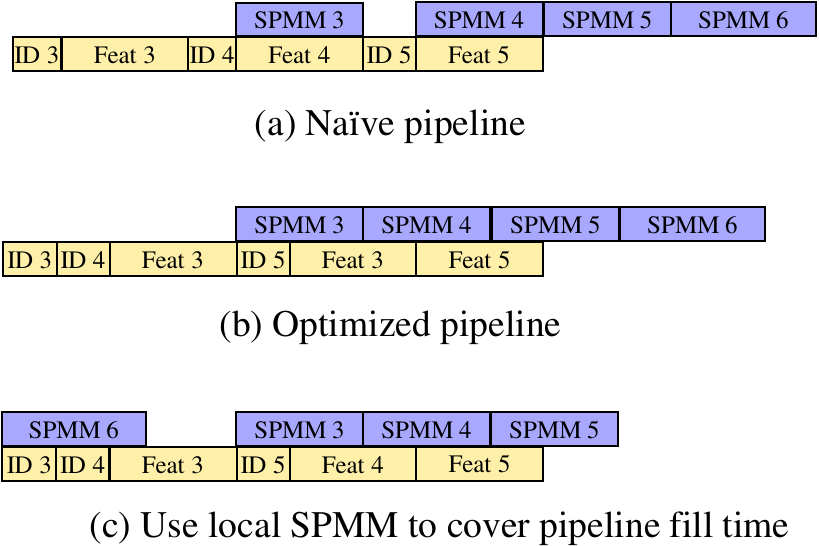}
    \caption{Pipelining the SPMM computation with four subgroups in rows 4-7 in Figure~\ref{fig:nz}. }
    \label{fig:pipeline}
\end{figure}

\textbf{Pipeline optimization.}
We organize the computation and communication of Figure~\ref{fig:nz} in the pipeline to hide the communication time. 
Figure~\ref{fig:pipeline}(a) shows an example of SPMM with four groups in rows 4-7 of ${\bf G_0}$, where each group is associated with two communications for column IDs and features except group 6.
We can schedule the groups in the pipeline so that the SPMM computation of the group overlaps with the communication of receiving features. For example, we first finish the communication for the column IDs of group 4. After that, we can start receiving the features for group 4 and sending the features to other machines while performing the SPMM computation of group 3. However, communicating the features depends on the results of ID communication, so we cannot start the SPMM before it completes, leading to the bubble between two SPMM computations.

We propose two reordering optimizations to reduce the communication cost in our pipelined strategy: 
(i) At the start of the primitive, we first communicate the IDs for groups 3 and 4 as shown in Figure~\ref{fig:pipeline}(b).
As a result, the SPMM computation of group 3 can overlap the communication of column IDs for group 5 and the features for group 4.
The communication for the features of the next group and the IDs of the group after the next group do not need synchronization. (ii) We can schedule the local SPMM (group 6) at the beginning to cover the pipeline fill time, as shown in Figure~\ref{fig:pipeline}(c).

\begin{figure}[ht]
    \centering
    \includegraphics[page=1,width=.7\columnwidth]{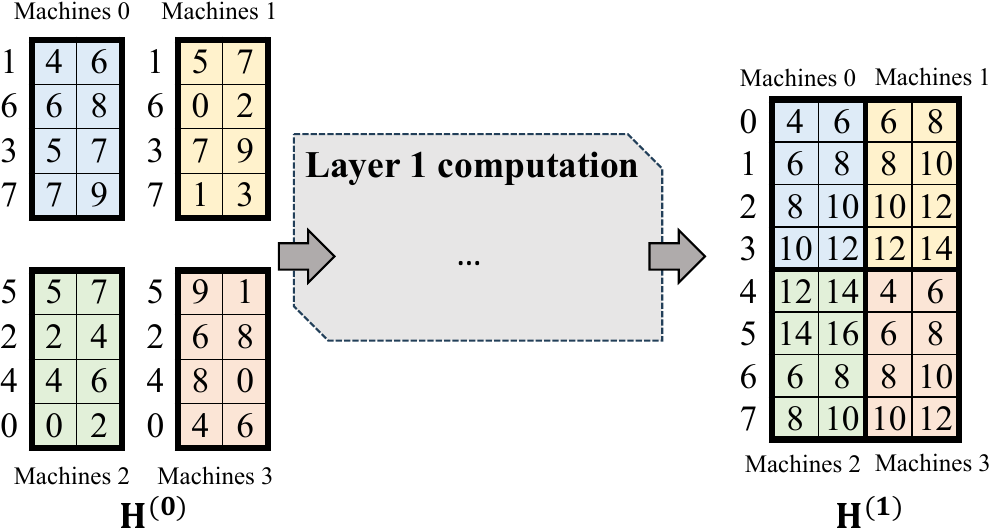}
    \caption{Random ordered feature tensor for layer 1 computation with an ordered output feature tensor. }
    \label{fig:system}
\end{figure}

\textbf{Fusing feature preparation with the first GNN primitive.} 
During GNN inference, we need to load the node features from the files. Of note, the feature files are not sorted based on IDs. Since {\name} partitions the features for scalability purposes, one can either let each machine scan all the feature files to obtain its own features or let each machine load part of the features and communicate for the correct feature distributions. When there are $M$ machines and $N$ nodes, the former approach incurs $O(M\cdot N)$ traffic on the file system, and the latter reduces the file system traffic by $M$ times and leads to $O(\frac{(M-1)N}{M})$ network traffic. Because the network has a larger aggregated bandwidth than the file system~\cite{efs}, we opt to let each machine load part of the features and then redistribute them.

{\name} goes further to avoid an extra redistribution cost as follows: 
(i) We build a table recording the location of each node feature on every machine.
In the example of Figure~\ref{fig:system}, machine 0 (blue) loads the features of nodes 1, 6, 3, and 7.
(ii) We let the machines that are supposed to hold a particular feature tile compute that tile in $\bf H^{(1)}$, so the residence of the first layer output, $\bf H^{(1)}$, aligns with the partition results. 
For example, for the sparse primitives in the first GNN layer, machines receive $\bf H^{(0)}[6]$ from machines 0 and 1 and receive $\bf H^{(1)}[6]$ from 2 and 3.

\section{Evaluation}\label{sec:eval}

\subsection{Experimental setup}
\begin{table}[ht]
\vspace{-.2in}
  \caption{Real-world graph datasets.}
  \vspace{-.1in}
  \resizebox{\columnwidth}{!}{
  {\scriptsize
  \begin{tabular}{cccc}
    \toprule
     Dataset & ogbn-products & social-spammer & ogbn-papers100M \\
    \midrule
    Nodes         & 2.4 M  & 5.6 M  & 111 M  \\
    Edges         & 123 M & 858 M & 1.6 B  \\
    \bottomrule
  \end{tabular}
  }
  }
    
    \label{tb:dataset}
    
\end{table}

\textbf{Datasets.} We conduct experiments on three real-world datasets as shown in Table~\ref{tb:dataset}. The \textit{ogbn-papers100M}~\cite{wang2020microsoft} is a citation graph whose nodes represent papers and edges are the citations. 
\textit{ogbn-products}~\cite{Bhatia16} depicts a product co-purchasing network, where nodes represent products sold on Amazon, and edges between two products indicate that they are purchased together.
The \textit{social-spammer}~\cite{fakhraei2015collective} dataset depicts a multi-relation social network with
legitimate users and spammers. 
Besides, we use synthetic datasets to evaluate scalability, generated using RMAT~\cite{chakrabarti2004r}, with the edge probabilities as $\{0.57, 0.19, 0.19, 0.05\}$, and the average degree as 20. 

\textbf{Models.}
We test the inference of 3-layer GCN and GAT. The hidden dimension of node features is set the same as the input feature dimension, which is 100 for ogbn-products and 128 for other datasets. The GAT model has 4 heads. We sample 50 neighbors for every GNN layer. 

\textbf{Baseline systems.}
We compare with the baseline system for {\name}'s GNN computation and graph construction.
In particular, we implement the GNN computation of DGI and SALIENT++ in DistDGL~\cite{zheng2020distdgl}. Note that we don't use P$^3$ as a baseline because it is not open-sourced.
For graph construction, {\name} is compared with DistDGL's built-in pipeline~\cite{dglpart}.

\textbf{System implementation}. 
We implement {\name} on top of DGL and PyTorch. Specifically, we leverage DGL for graph operations and PyTorch's distributed package for communication. The experiments are run with PyTorch 2.0 and DGL 1.1. The system is deployed on up to 16 AWS R5.16xlarge EC2 instances with Intel Xeon Platinum 8175 and 768 GB memory. Instances are connected via 25 Gbps Ethernet.

\subsection{{\name} vs State-of-the-art}

\begin{figure}[ht]
    \centering
    \includegraphics[width=.8\columnwidth]{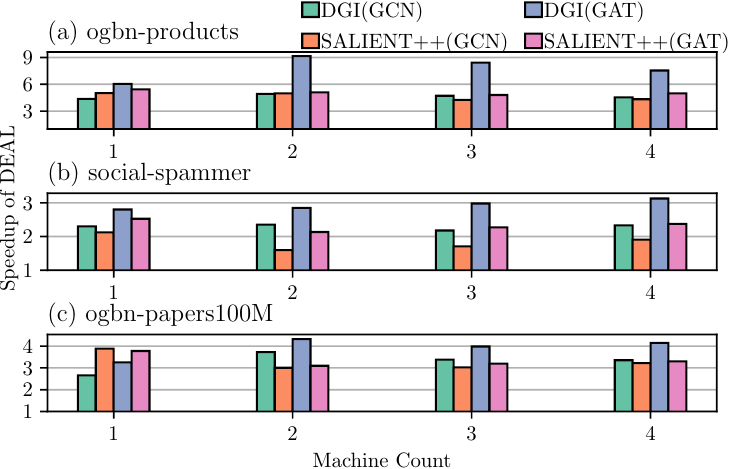}
    \caption{The speedup of {\name} over DGI and SALIENT++. 
    }
    \label{fig:eva:baseline}
\end{figure}

\textbf{{\name} v.s SOTA.}
Figure~\ref{fig:eva:baseline} shows the speedup of {\name} over DGI and SALIENT++ across three datasets and two models. For GCN, {\name} achieves $4.64\times$, $2.28\times$, and $3.25\times$ speedup over DGI, and $4.36\times$, $1.82\times$, and $3.26\times$ speedups over SALIENT++, respectively for the three datasets. For GAT, over the three datasets, respectively, {\name} enjoys $7.70\times$, $2.93\times$, and $3.90\times$ speedups against DGI, and $3.07\times$, $1.32\times$, and $2.32\times$ speedup against SALIENT++. Regarding the trends on datasets, as the graph grows larger (ogbn-papers100M) and sparser (ogbn-products), DGI suffers from decreased sharing ratios as such graphs are harder for DGI to obtain common neighbors, and SALIENT++ experiences increased overhead from building and maintaining its cache.

Regarding the model trends, {\name} achieves higher speedup on GAT when compared with DGI because GAT contains more primitives, benefiting more from better exploited sharing. In contrast, {\name} exhibits higher speedups for GCN when compared against SALIENT++, because its GCN computation is dominated by feature communication, which {\name} significantly improves. 
{\name} keeps similar speedups when increasing the number of machines.
The reason is that our sampling would offer better speedups, but it results in more communication. These two effects are roughly comparable, thus leading to maintained speedups.

\begin{table}[ht]
  \caption{The sharing ratio of different approaches.}
  \vspace{-.1in}
  \resizebox{\columnwidth}{!}{
    \begin{tabular}{cccc}
      \toprule
        & ogbn-products & social-spammer & ogbn-papers100M \\
      \midrule
      DGI~\cite{yin2023dgi}   & 60.1\%       & 87.0\% & 63.9\%  \\
      P$^3$~\cite{gandhi2021p3}    & 33.3\%       & 46.1\% & 28.6\% \\
      SALIENT++~\cite{kaler2023communication}   & 66.4\%       & 77.9\% & 70.3\% \\
      \bottomrule
    \end{tabular}
  }

  \label{tab:eva:share}
  
\end{table}

\textbf{Sharing ratio.} 
Table~\ref{tab:eva:share} shows the sharing ratio of different approaches. Across the three datasets, DGI, P$^3$, and SALIENT++ achieve an average sharing ratio of $70.3\%$, $36\%$, and $71.5\%$, respectively. Although P$^3$ can leverage all sharing in the outermost hop, the outermost hop alone only contributes limited sharings, so P$^3$ has the lowest overall sharing ratio. When comparing the different datasets, the three approaches $51.0\%$, $67.8\%$, and $50.4\%$, respectively. Notably, there is an inverse relationship between the achieved sharing ratio and the speedup of {\name}. SALIENT++ has a higher sharing ratio than DGI, but its cache maintenance overhead slows it.

\begin{table}[ht]
\caption{The test accuracy on ogbn-products.}
  \vspace{-.1in}
  \resizebox{\columnwidth}{!}{\scriptsize
    \begin{tabular}{cccc}
      \toprule
        Model & Full neighbor & SALIENT++ & Ours \\
      \midrule
      GCN   & 76.9\% ($\pm$0.29\%)       & 76.9\% ($\pm$0.46\%) & 76.9\% ($\pm$0.43\%)  \\
      GAT   & 79.4\% ($\pm$0.12\%)        & 79.3\% ($\pm$0.63\%)  & 79.2\% ($\pm$0.82\%) \\
      \bottomrule
    \end{tabular}
  }
  \label{tab:acc}
\end{table}

\textbf{Accuracy study.}
We evaluate the accuracy of {\name} on the ogbn-products in Table~\ref{tab:acc}. {\name} reuses the same sampled 1-hop ego networks for different nodes, which is slightly different from the conventional mini-batch inference~\cite{kaler2023communication, zheng2020distdgl}. However, our results show that {\name} achieves similar or the same accuracy as sampling-based method (i.e., SALIENT++). Particularly, this study compares the layer-by-layer inference in {\name} with the full neighbor inference and the mini-batch inference in SALIENT++. We trained two 3-layer GCN and GAT models with sampling fanout as 10. 
{\name} achieves the same accuracy for GCN and similar accuracy for GAT when compared to SALIENT++ and full neighbor-based approach.

\subsection{Scalability}

\begin{figure}[ht]
\vspace{-.2in}
\centering
\begin{tabular}{cc}  
    \subfloat[Weak scaling.]{
        \hspace{-.2in}
    \includegraphics[width=0.4\columnwidth]{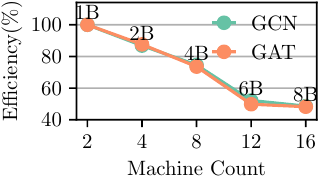}
    }
    &
    \subfloat[obgn-products.]{
        \hspace{-.2in} 
    \includegraphics[width=0.4\columnwidth]{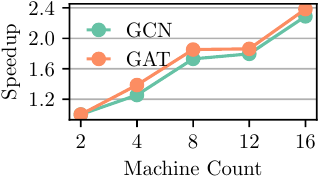}
    }
    \\
    \subfloat[social-spammer.]{
    \vspace{-.2in}
        \hspace{-.2in} 
    \includegraphics[width=0.4\columnwidth]{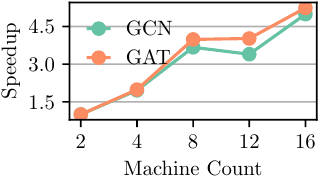}
    }
    &
    \subfloat[ogbn-papers100M.]{
        \hspace{-.2in} 
    \includegraphics[width=0.4\columnwidth]{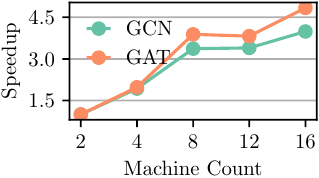}
    }
\end{tabular}
\caption{Scalability test of {\name}: (a) Weak scaling on synthetic data. (b-d) Strong scaling for (b) ogbn-products, (c) social-spammer, and (d) ogbn-papers100M.}
\label{fig:eva:scale}
\end{figure}

We evaluate the scalability of {\name} using synthetic datasets. We use processed edges per second per machine to represent the system efficiency. Figure~\ref{fig:eva:scale}(a) shows the weak scaling of the GNN computation
We run graphs of different scales on different cluster sizes. For example, we run a graph with 1B edges on 2 machines and a graph with 8B edges on 16 machines. When scaled to 16 machines, {\name} retains $48.2\%$ and $47.9\%$ efficiency compared with using 2 machines for GCN and GAT, respectively.  

Figure~\ref{fig:eva:scale}(b), (c), and (d) shows the strong scalability from 2 machines to 16 machines on real-world datasets. 
When scaled to 16 machines, {\name} retains $2.28\times$, $4.98\times$, and $3.98\times$ for GCN, and $2.38\times$, $5.32\times$, and $4.83\times$ for GAT. Compared with GCN, GAT has better scalability because it has more GEMM primitives. When graphs grow larger, the scalability of {\name} is better because the fixed overhead such as communication latency becomes insignificant.

\subsection{Distributed primitive evaluation}
\subsubsection{GEMM}

\begin{figure}[ht]
\centering
\begin{tabular}{cc}  
    \subfloat[Dimension = 256.]{
        \hspace{-.2in} 
    \includegraphics[width=0.42\columnwidth]{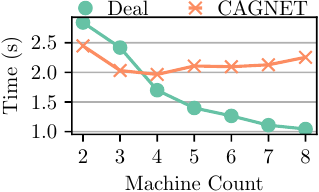}
    \label{fig:eva:gemm_256}
    }
    &
    \subfloat[Dimension = 1024.]{
        \hspace{-.2in}
    \includegraphics[width=0.42\columnwidth]{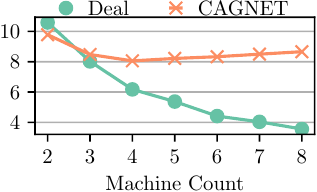}
    \label{fig:eva:gemm_1024}
    }
\end{tabular}
% \vspace{-.1in}
\caption{The evaluation of our distributed GEMM for ogbn-products with hidden dimensions 256 and 1024.}
\label{fig:eva:GEMM}
\end{figure}

Figure~\ref{fig:eva:GEMM} evaluates the distributed GEMM algorithm ({\name} vs. CAGNET) on ogbn-products for two sizes of hidden dimensions. As GEMM performance is graph-structure independent, we restrict our results to one dataset. {\name}'s distributed GEMM approach demonstrates substantial scalability, with average speedups of 1.97$\times$ and 2.97$\times$ when using 4 and 8 machines, respectively, compared to the 2-machine baseline. While {\name} experiences noticeable overhead of adjusting the memory layout to accommodate the communication library for 2 and 3 machines, this overhead becomes trivial when the machine number is large. In contrast, CAGNET's GEMM exhibits poorer scalability due to increased communication overhead with more machines. 
Overall, benefiting from reduced communication, our method significantly outperforms CAGNET, achieving average speedups of 1.52$\times$ and 1.47$\times$ across different machine counts. The speedup increases with more machines used.

\subsubsection{SPMM}

\begin{figure}[ht]
    \centering
    \includegraphics[width=0.8\columnwidth]{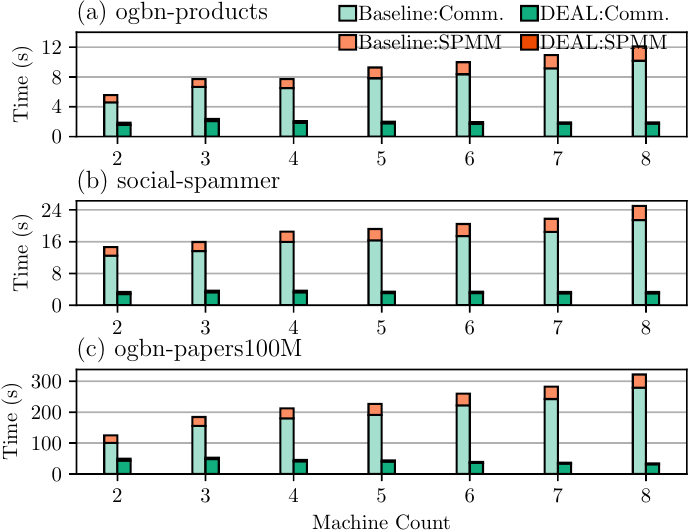}
    \caption{The performance comparison of baseline graph exchange SPMM and {\name}'s feature exchange SPMM.}
    \label{fig:eva:SPMM}
\end{figure}

Figure~\ref{fig:eva:SPMM} shows the performance of SPMM, evaluating (i) exchanging graph structure as the baseline and (ii) exchanging features in {\name}. When comparing the two options, {\name} achieves $4.30\times$, $5.28\times$, and $5.29\times$ speedups for three datasets, respectively. The speedup contains two parts, i.e., communication and SPMM computation. For communication, the reduced communication of {\name} enjoys $4.15\times$, $5.30\times$, and $4.86\times$ speedups over the baseline, respectively. 
For SPMM computation, 
{\name} delivers $6.14\times$, $7.21\times$, and $8.78\times$ speedups over the baseline, respectively.

Moreover, the scalability of these approaches diverges significantly. As the number of machines increases from 2 to 8, baseline shows decreased performance, becoming $2.27\times$, $1.52\times$, and $2.49\times$ slower, respectively. In contrast, {\name} achieves $1.21\times$, $1.08\times$, and $1.52\times$ speedup, showcasing its superior scalability over the baseline. The reason is that the size of the sparse matrix sparse matrix tile does not reduce linearly as the number of partitions increases. Therefore, the baseline experiences a larger communication size to exchange the spare matrix tile when \# partitions increases.

\subsubsection{SDDMM}

\begin{figure}[ht]
% \vspace{-.2in}
    \centering
    \includegraphics[width=\columnwidth]{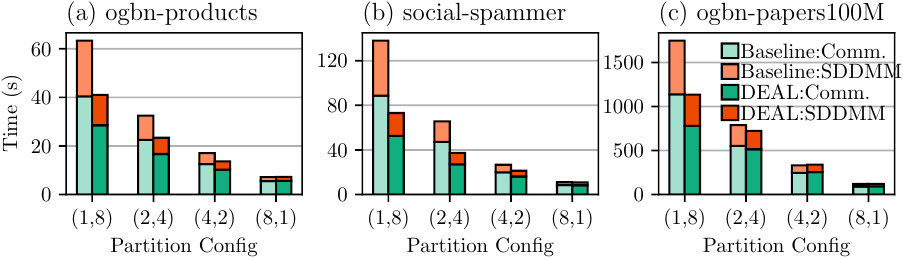}
    \caption{The performance of SDDMM across varying partitioning configurations, denoted as (\#graph partitions, \#feature partitions). For example, (1,8) means 1 graph partition and 8 feature partitions 
    % \color{red}{Xiang: what does (1,8),(2,4) mean?}
    }
    \label{fig:eva:SDDMM}
\end{figure}

Figure~\ref{fig:eva:SDDMM} evaluates the SDDMM under various partitioning configurations. In SDDMM, machines communicate with other machines storing the different graph partitions and feature partitions. Therefore, the total communication is a combined effect of graph partition and feature partitions. 
We used fixed eight machines and varied the number of graph and feature partitions to assess their impacts on communication and computation times. The two approaches examined are (i) duplicating computation across partitions (baseline) and (ii) splitting non-zeros among partitions ({\name}). As we increase feature partitions from one to eight, {\name} demonstrates speedups of $1.65\times$, $1.38\times$, $1.15\times$, and $1.00\times$. Notably, both approaches are equivalent with a single feature partition (hence $1.00\times$ for the last case).
Regarding communication efficiency, as shown as light green bars and deep green bars for baseline and {\name} respectively, {\name} yields speedups of $1.32\times$, $1.53\times$, and $1.15\times$, respectively, across the partition configurations. Moreover, the exploitation of computation parallelism under {\name} results in speedups of $1.54\times$, $1.78\times$, and $1.24\times$, respectively, when the number of graph partitions increases. 
Dataset comparison reveals that denser graphs, such as those from the social-spammer dataset, benefit more from computational speedup, while larger and sparser graphs, like ogbn-papers100M, see reduced speedup primarily due to the communication overhead in aggregating edge features computed across machines.

\subsection{Study system implementation optimizations}

\begin{figure}[ht]
\centering
\begin{tabular}{c}  
    \subfloat[SPMM.]{
        \hspace{-.2in}
    \includegraphics[width=.9\columnwidth]{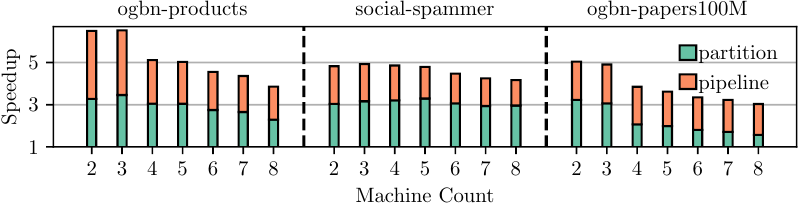}
    \label{fig:eva:SPMM_system}
    }
    \\
    \subfloat[SDDMM.]{
    \vspace{-.2in}
        \hspace{-.2in} 
    \includegraphics[width=.9\columnwidth]{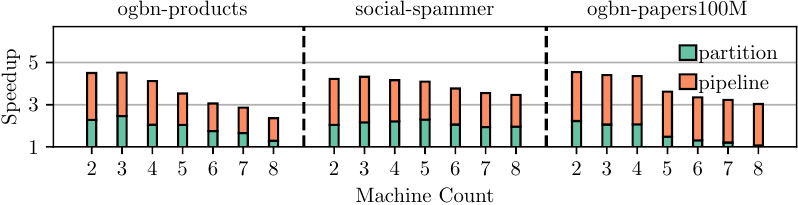}
    \label{fig:eva:SDDMM_system}
    }
\end{tabular}
\caption{The speedup of {\name} with partitioned communication and pipelining.}
\label{fig:eva:system}
\end{figure}

\textbf{Partitioned communication and pipeline optimization.}
Figure~\ref{fig:eva:system} depicts the speedup achieved by the sparse primitives of {\name} through two optimizations: partitioned communication and pipelining. Across datasets and machine counts, the partitioned communication yields the average speedups of $2.90\times$, $3.09\times$, and $2.15\times$ for SPMM, and $1.89\times$, $2.09\times$, and $1.57\times$ for SDDMM, respectively. Denser graphs with more non-zeros per column benefit more due to efficient communication merging, leading to the highest speedup for the dense social-spammer dataset and the lowest gain on the sparser ogbn-papers100M. The speedup decreases with more machines as the ratio of redundant communication is reduced. Compared with SPMM, the speedup of SDDMM is smaller because we assign the non-zeros to different machines row-wise, reducing the number of non-zeros in each group. 
Subsequently, applying pipelining further boosts performance on average by $1.50\times$, $1.65\times$, and $1.47\times$ for SPMM, and  $1.82\times$, $2.15\times$, and $1.90\times$ for SDDMM. The dense graphs enjoy more speedup due to reduced communication overhead. Likewise, the SDDMM achieves higher speedup because of more communication operations per group. 
Cumulatively, our optimizations achieve overall speedups of $4.41\times$, $4.74\times$, and $3.61\times$ on SPMM, and  $3.72\times$, $4.24\times$, and $3.50\times$ on SDDMM by combining partitioned communication and pipelining across the respective datasets, underscoring the compounded benefit.

\begin{figure}[ht]
    \centering
    \includegraphics[width=.8\columnwidth]{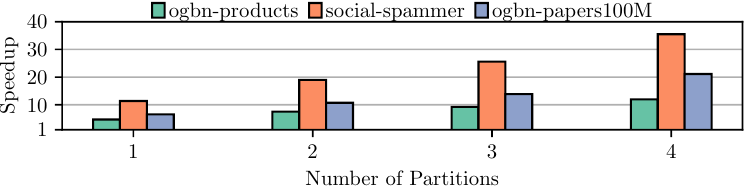}
    \caption{The speedup of graph construction over DistDGL.}
    \label{fig:eva:construction}
    % \vspace{-.2in}
\end{figure}

\textbf{Graph construction.}
Figure~\ref{fig:eva:construction} illustrates the speedup of our graph construction over DistDGL. {\name} achieves $7.92\times$, $21.05\times$, and $11.99\times$ speedup on average across the evaluated datasets, respectively. 
{\name} exhibits higher speedup on graphs with more edges because DistDGL can only process the edge list using one machine while {\name} fully distributes the construction. 
Furthermore, leveraging 4 machines in {\name} results in $2.54\times$, $3.11\times$, and $3.21\times$ speedup when compared to a single machine for the respective datasets. This scaling efficiency is particularly pronounced for larger graphs. 

\begin{figure}[ht]
    \centering
    \includegraphics[width=0.8\columnwidth]{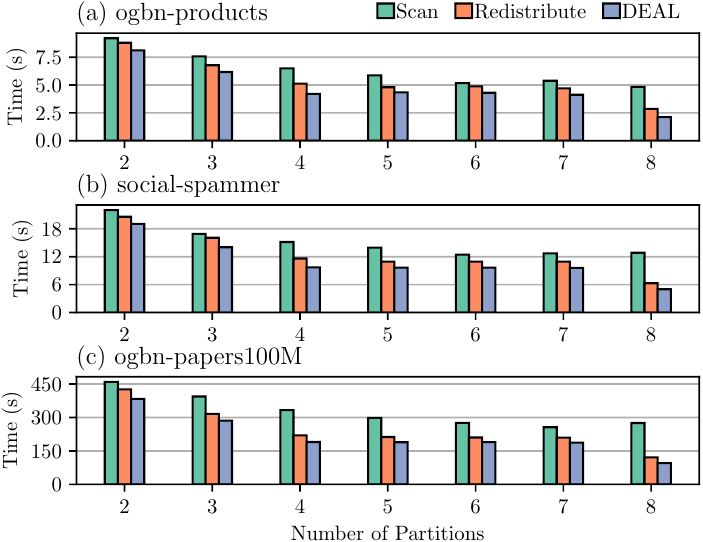}
    % \vspace{-.2in}
    \caption{Impacts on {\name} for feature preparation.
    % \vspace{-.2in}
    }
    \label{fig:eva:redist}
\end{figure}

\textbf{Feature preparation.} 
Figure~\ref{fig:eva:redist} evaluates the fusing feature preparation with the first GNN primitive across three distinct datasets. Notably, when compared to the baseline scan-through loading method, the feature redistribution design achieves a speedup of 1.20$\times$, 1.26$\times$, and 1.39$\times$ on average for these datasets, respectively, across varying machine counts. Furthermore, {\name}'s communication-free method yields additional 1.15$\times$, 1.15$\times$, and 1.14$\times$ speedups, respectively. As we scale up the number of machines, the baseline time remains unchanged because the file system is the bottleneck.
In contrast, when the machine count increases from 2 to 8, the redistribution approach achieves a speedup of 3.27$\times$ over the baseline using 2 machines. {\name} further achieves 3.88$\times$ over 2-machines baseline, underscoring the benefits of communication reduction.

\section{Related work}\label{sec:related}

GNN research has proliferated recently. GNN applications~\cite{hamidi2023learning,sun2020benchmarking,suzuki2023clustered}, algorithms~\cite{zhu2021graph,chen2024view,khan2023interpretability,jiang2022fast,tao2022cross,wang2023efficient,wu2023certified,FENG2023119617}, models~\cite{kipf2016semi,hamilton2017inductive,layne2023temporal}, systems~\cite{chen2023tango,zheng2020distdgl,huang2024wisegraph, chen2023compressgraph}, hardware~\cite{FINGERS,yu2023race,lyu2022efficient}, among many others~\cite{zheng2022distributed,zheng2020dglke,calvanese2023conceptually,Graph2020Amirali}. We refer the readers to a handful of GNN surveys for a more comprehensive landscape of GNN research~\cite{abadal2021computing}. This work mainly focuses on two related subjects: distributed GNN and GNN inference acceleration. 

\textbf{Distributed GNN computation} can be categorized through two avenues: {(i) ego network-centric distribution} and {(ii) full graph-centric distribution.} 
Below, we discuss them separately. 

{(i) Ego network-centric distribution} treats the ego network as a first-class citizen, and distribution is achieved centering around each ego network entity. PyG~\cite{fey2019fast}, DGL~\cite{wang2019dgl}, AliGraph~\cite{yang2019aligraph}, AGL~\cite{zhang2020agl}, DistDGL~\cite{zheng2020distdgl}, P$^3$~\cite{gandhi2021p3}, FlexGraph~\cite{wang2021flexgraph}, Betty~\cite{yang2023betty}, SALIENT++~\cite{kaler2022accelerating}, and PaGraph~\cite{lin2020pagraph} belong to this category.
We discuss some representative projects: 
P$^3$~\cite{gandhi2021p3} introduces the hybrid parallelism to address the redundancy. All machines first exchange the ego networks and collectively compute the features of all nodes in the first layers. Then, the results are communicated, and every machine continues for the rest of the layers of each ego network. 
FlexGraph~\cite{wang2021flexgraph} dynamically migrates the ego networks among the machines to balance the workload and minimize the communication cost. 
Betty~\cite{yang2023betty} partitions the multi-layer bipartite graph built by a batch of ego networks. The goal is that each ego network owns one graph partition, and the inter-partition communication is reduced to mitigate redundancy. 
SALIENT++~\cite{kaler2022accelerating} and PaGraph~\cite{bai2021efficient} focus on caching features of hub nodes, which are often included in multiple ego networks, to eliminate the need for repeated communication. 

{\name} is fundamentally distinct from the aforementioned paradigm. Particularly, {\name} breaks all ego networks into 1-hop samples and computes all samples of the same layer together with distributed primitives. This process continues layer-by-layer to arrive at the final embeddings for all nodes. This concerted effort eliminates redundancy and offers rich pipelining opportunities. 

{(ii) Full graph-centric distribution} simply partitions the graph for distributed GNN.
NeutronStar~\cite{wang2022neutronstar}, Sancus~\cite{peng2022sancus}, NeuGraph~\cite{ma2019neugraph} DistGNN~\cite{md2021distgnn}, DGCL~\cite{cai2021dgcl}, and Dorylus~\cite{thorpe2021dorylus} fall in this category.  
Particularly, DGCL~\cite{cai2021dgcl} introduces a novel communication scheduling approach that considers both network topology and GNN computation dependencies to reduce communication costs.
NeutronStar~\cite{wang2022neutronstar} opts for an adaptive solution between recomputation and caching to reduce communication costs. 
The benefits are pronounced when nodes have few dependencies and larger hidden layer sizes, where computation overhead is less than the communication overhead.
NeuGraph~\cite{ma2019neugraph} distributes tiles of the adjacency matrix across multiple GPUs. Each GPU calculates partial results and then employs an all-reduce operation for complete results aggregation.
Besides, DistGNN~\cite{md2021distgnn} and Sancus~\cite{peng2022sancus} delay the communication of sparse primitives and proceed with partial aggregated results. This approach, while effective, can lead to accuracy losses.

{\name} is different as follows: {\name} partitions all the participating tensors during GNN distribution, including the sparse graph tensor, and node and edge feature tensors. We prioritize the feature tensors due to their superior size. In contrast, the aforementioned projects only focus on graph partitions.

\textbf{Distributed GNN primitives.} Existing work on optimizing distributed primitives for GNNs focuses on reducing communication overhead~\cite{koanantakool2016communication, tripathy2020reducing, kurt2023communication, selvitopi2021distributed, zhang2023unfairness}. \cite{selvitopi2021distributed} and \cite{koanantakool2016communication} proposes novel partition and associated communication algorithms to optimize individual primitives.
MGG~\cite{wang2023mgg} leverages the sparsity to overlap the communication and computation within a GPU kernel. Techniques like CAGNET~\cite{tripathy2020reducing} optimize work distribution based on the GNN computation flow, and RDM~\cite{kurt2023communication} redistributes matrices to accommodate different primitives. However, these efforts lack support for diverse GNN models~\cite{velivckovic2018graph, hamilton2017inductive} and are coupled with specific model structures, limiting their applicability.

\textbf{GNN inference acceleration.} A separate line of research has focused on optimizing GNN inference from various perspectives~\cite{he2022coldguess,hosseini2023exploring,lyu2022efficient,abadal2021computing,tan2023quiver,yang2024gmorph}. Work by~\cite{zhou2021accelerating} taps into model pruning to diminish the hidden dimension of node representations. Similarly, ~\cite{auten2020hardware} devises hardware architectures tailored to efficiently manage the irregular data movement inherent to GNNs. 
HAG~\cite{jia2020hag} is proposed to reduce the redundancy computation in neighbor aggregation by combining the common neighbors of different nodes. HAG can reduce the total aggregation operations, but searching for the neighbor combination is time-consuming.

\section{Conclusion}\label{sec:conclude}

{\name} introduces distributed end-to-end GNN inference at scale for all nodes. Particularly, {\name} makes three major contributions. First, {\name} introduces a lightweight partitioning strategy for end-to-end inference. Second, {\name} designs the distributed GNN primitives to address partitioned graphs and features communication and memory consumption issues. Third, {\name} implements partitioning and scheduling mechanisms to reduce communication costs further and enable pipelining-based optimizations.
With {\name}, the end-to-end inference time on real-world benchmark datasets is reduced up to $7.70\times$ and the graph construction time is reduced up to $21.05\times$, compared to the state-of-the-art.

\bibliographystyle{unsrt}
\bibliography{ref}

\end{document}